the past. Although agriculture may have arisen there over 6500 years ago, highland New Guinea societies are still relatively egalitarian and characterized by "big men," whose influence is largely persuasive and consensual. The evidence for early agriculture from highland New Guinea signifies the potential diversity of prehistoric trajectories after the inception of agriculture and challenges unilinear, often teleological, interpretations of human prehistory.

The authors are indebted to J. Golson for his generous assistance and encouragement during every stage of the work documented here and for unfettered access to the site archive for the 1970s investigations. J. Golson, G. Hope, M. Spriggs, S. Blau, and anonymous reviewers are thanked for their comments on a draft manuscript. T.P.D. acknowledges scholarships and a fieldwork grant from the Australian National University (ANU) and grants from the Australia-Pacific Science Foundation (awarded to J. Golson in 1998 and 1999) and the Pacific Islands Development Program (1999). S.G.H. completed this work while on an Australian Research Council QEII Fellowship. C.L. acknowledges a scholarship from Southern Cross University and grants from the Pacific Biological Foundation (2002) and Australian Museum. The Centre for Archaeological Research at ANU, Australian Nuclear Science and Technology Organisation, and Research School of Earth Sciences, ANU (courtesy of J. Chappell) provided radiometric age determinations. Thanks are due to K. Dancey, R. Patat, and A. and C. Rohn for their assistance with the graphics. Current studies were made possible with the permission and assistance of the Papua New Guinea National Museum and Art Gallery, National Research Institute, Western Highlands Provincial Government, and the Kawelka at Kuk.

**Supporting Online Material**
www.sciencemag.org/cgi/content/full/1085255/DC1
Tables S1 to S3
References

2 April 2003; accepted 16 May 2003
Published online 19 June 2003;
10.1126/science.1085255
Include this information when citing this paper.

# REPORTS

# A Young White Dwarf Companion to Pulsar B1620-26: Evidence for Early Planet Formation

Steinn Sigurdsson,[1]* Harvey B. Richer,[2] Brad M. Hansen,[3] Ingrid H. Stairs,[2] Stephen E. Thorsett[4]

The pulsar B1620-26 has two companions, one of stellar mass and one of planetary mass. We detected the stellar companion with the use of Hubble Space Telescope observations. The color and magnitude of the stellar companion indicate that it is an undermassive white dwarf (0.34 ± 0.04 solar mass) of age $480 \times 10^6 \pm 140 \times 10^6$ years. This places a constraint on the recent history of this triple system and supports a scenario in which the current configuration arose through a dynamical exchange interaction in the cluster core. This implies that planets may be relatively common in low-metallicity globular clusters and that planet formation is more widespread and has happened earlier than previously believed.

Messier 4 (M4 equals NGC 6121 and GC 1620−264) is a medium mass [$\sim 10^5$ solar mass ($M_\odot$)] globular cluster and the one closest to the Sun. It has a moderately dense ($\rho_0 \approx 3 \times 10^4 \, M_\odot \, \text{pc}^{-3}$) core. The metal content of the cluster is 5% that of the Sun, with little variation in composition or age between different member stars. The cluster has a substantial population of white dwarfs (stellar remnants which have exhausted their nuclear fuel), recently detected in deep Hubble Space Telescope (HST) observations (*1*, *2*), that have been used to determine an age for the cluster of $12.7 \times 10^9 \pm 0.35 \times$ $10^9$ years. Furthermore, M4 contains the binary radio pulsar PSR B1620−26 (*3*, *4*), a recycled millisecond pulsar with a $P = 11$ ms rotation period and a companion in a low eccentricity ($e = 0.025$) orbit with an orbital period of 191 days. For an assumed pulsar mass of 1.35 $M_\odot$, radio timing observations constrain the companion mass to be $M_c = 0.28 \, M_\odot/(\sin i)$, where $i$ is the unknown inclination of the binary orbital plane to the line of sight (*5*, *6*). The pulsar also possesses an anomalously large second time derivative of the rotational period ($\ddot{P}$) (*7*, *8*), seven orders of magnitude larger than that expected from the intrinsic pulsar spin-down and of the wrong sign. When discovered, the pulsar had a characteristic spin-down time

[1]525 Davey Laboratory, Department of Astronomy, Pennsylvania State University, University Park, PA 16802, USA. [2]Department of Physics and Astronomy, University of British Columbia, 6224 Agricultural Road, Vancouver, British Columbia V6T 1Z1, Canada. [3]Department of Physics and Astronomy and Institute of Geology and Planetary Physics, University of California at Los Angeles, Math-Sciences 8971, Los Angeles, CA 90095–1562, USA. [4]Department of Astronomy and Astrophysics, University of California, Santa Cruz, CA 95064, USA.

*To whom correspondence should be addressed. E-mail: steinn@astro.psu.edu

R E P O R T S

scale (or "age") $\tau_c = P/2\dot{P} = 2.2 \times 10^8$ years (where $\ddot{P}$ is the time derivative of the rotation period), similar to that of other globular cluster pulsars, though the time scale for the period derivative to change, $\tau_p = \dot{P}/\ddot{P}$, was only about 10 years. In the last few years, the period derivative has changed sign, so the pulsar period appears to be shortening.

Further radio observations revealed the third and fourth time derivatives of the pulse period and also a secular change in the projected semimajor axis of the inner binary (5, 6). The accumulated evidence points toward a model in which these timing residuals are caused by the time-varying acceleration of the pulsar due to the gravitational influence of a second companion. This can be either a planetary mass companion in a moderate or low eccentricity orbit (7–10) or a stellar mass companion in a wide, high-eccentricity orbit (7, 8, 11, 12). Further dynamical modeling concluded that the companion was probably a substellar mass object. An important success of this model was an explanation of the eccentricity of the inner orbit as the result of Kozai pumping (13, 14).

On the basis of these constraints, a model was introduced to explain the origin of the current configuration (9, 10, 15). The planet was initially in orbit around the main-sequence star progenitor of what is now the white-dwarf companion. This system underwent an exchange interaction with a neutron star binary, with the main-sequence star progenitor replacing the original white-dwarf companion to the neutron star. The end results were a new, eccentric orbit for the main-sequence star progenitor with a semimajor axis similar to or larger than that of the original binary and the displacement of the planet onto a wide, circumbinary orbit. The binding energy of the new binary increased because the new stellar companion was more massive, and the resulting energy release led to a recoil of both the ejected white dwarf and the new binary. This recoil displaced the binary from the core. The post-main-sequence evolution of the progenitor, over about 1 billion years, led to mass transfer and spin-up of the neutron star to its current spin period. Although quite exotic, this scenario is testable because it makes a clear prediction, namely that the current configuration was established relatively recently ($<10^9$ years). In particular, the wide orbit of the planet makes it vulnerable to disruption in the denser regions of the cluster, suggesting that the system has not yet returned to the core after the interaction. An ejected binary will sink back to the core on a time scale on the order of the cluster half-mass relaxation time, $\sim10^9$ years. Because the scenario outlined above requires the white dwarf to have lost its envelope since the exchange interaction, this provides a prediction: The white dwarf companion should be young, undermassive, and relatively bright. An alternative scenario has the pulsar binary capture a planet from another passing main-sequence star through dynamical exchange after the white dwarf forms (16). This scenario implies the white dwarf would probably be old.

A multifield, multiepoch WFPC2 HST-imaging study of this cluster (1, 17) was carried out in 1995 (GO-5461) and 2001 (GO-8679). We also secured archival observations taken in 2000 (GO-8153). The fields observed within the cluster were located at about 1, 2, and 6 core radii ($r_c = 50''$). The pulsar PSR B1620−26, located in M4, had J2000 coordinates for the right ascension (RA) of 16h 23m 38.2218 and for the declination (DEC) of −26° 31′ 53.769″ on julian date (JD) 2,448,725.5 (5), placing it near the edge of the inner field of these programs, about 46″ (about 1 $r_c$) from the cluster center.

Astrometry on the HST charge-coupled device chips was carried out with the use of the STSDAS task METRIC. Relative positions are reasonably accurate with this procedure (about 0.1″); however, our estimated uncertainty in the absolute positions of well-measured stars is ±0.7″ (18). We estimated the location of the pulsar in the three images in each bandpass (Fig. 1). Within a circle of radius 0.7″, only one star of an apparent magnitude $V = 24.0$ was observed, and we could measure objects 20 times fainter in these images. From an examination of the images in the various bandpasses, it is clear that the object within the circle is very blue (Table 1). The only other potential pulsar companion is the star at the left edge of the error circle. This star has a much redder color.

The multiple epoch observations allow us to separate cluster members from inner halo stars (the principal background contaminant for $V < 27$) with the use of the cluster proper motion (17, 19). The pulsar companion candidate is clearly a white dwarf and lies above the sequence delineated by the majority of the white dwarfs in the cluster (Fig. 2). The only other potential candidate (on the edge of the error circle) is a cluster main-sequence star with a mass of about 0.45 $M_\odot$ and is probably the object identified in previous, shallower, ground-based studies (20, 21). For completeness, we also examined the images obtained by Bailyn with HST in 1999 (GO-6166). In these short-exposure images at the pulsar position, there is a very faint object (formally below the detection limit) that is consistent with the position of the white dwarf we observe. The probability of getting a white dwarf in the error circle indicated on the images simply by chance is 0.6% (22).

The location of the white dwarf above and redward of the main cooling sequence is consistent with the mass inferred from pulsar timing and smaller than the canonical white dwarf mass $\sim0.5$ to 0.6 $M_\odot$. Such low-mass white dwarfs arise from the truncation of stellar evolution in close binaries (23, 24). The uncertainty in the extinction in the direc-tion of M4, and hence the absolute magnitude of the white dwarf, is ~0.1 magnitudes, limiting our constraint on the white dwarf mass to $M_{wd} = 0.34 \pm 0.04\ M_\odot$. The age of the white dwarf, taking into account the uncertainties in extinction and mass, is $4.8 \times 10^8 \pm 1.4 \times 10^8$ years, assuming a helium-core white dwarf with a hydrogen envelope of mass fraction $10^{-4}$ and with the use of the most recent models of Hansen and Phinney (25, 26). This confirms the proposed evolutionary scenario: The pulsar companion is indeed undermassive and younger than the time scale for an ejected system to return to the cluster core.

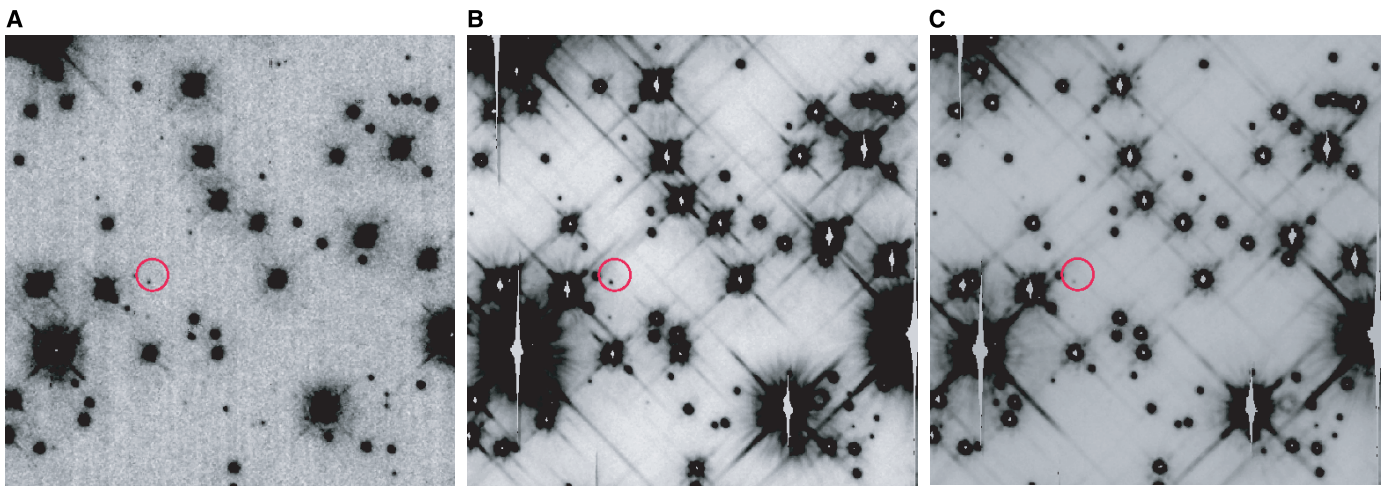

**Fig. 1.** (**A** to **C**) Hubble Space Telescope images of the field where the pulsar is located. The position of the pulsar is indicated by the center of the circle, which has a radius of 0.7″. The three images are the U (F336W), V (F555W), and I (F814W) bandpasses, which are wide-band filters centered on 336 nm, 555 nm, and 814 nm, respectively.

The observed mass function of the radio pulsar orbit, our new white dwarf mass, and the assumption that the pulsar mass is 1.35 $M_\odot$ allow an estimate of the orbital inclination of the inner binary to the line of sight: $55^{+14}_{-8}°$. This is consistent with the inclination ($40° \pm 12°$) predicted from models of the perturbation of the inner orbit by the gravitational force due to the planet (5, 6) and places the mass and semimajor axis of the planet at the lower end of their respective allowed ranges (5, 6, 13). The semimajor axis is about 23 astronomical units (AU) and the inferred planetary mass is ~2.5 $\pm$ 1 $M_{Jupiter}$, consistent with the range of masses found for Jovian planets in the solar neighborhood and inconsistent with typical brown dwarf masses.

The neutron star was probably originally a member of a binary with a high-mass white dwarf companion ($M_{WD}$ ~0.5 to 0.7 $M_\odot$) (27) and a semimajor axis of about 0.1 to 0.3 AU, analogous to the field pulsar binaries PSR J0621+1002 and PSR J1022+1001 (28). Such a binary sinks to the core of the cluster by virtue of its above-average mass. In the dense environment of the core, it encounters other stars, most probably stars with masses at or near the turn-off mass of the cluster main sequence (9, 10). The resultant exchange and recoil ejects the system from the dense core to the lower density regions of the cluster.

Within this scenario, the planet has its origin in a standard ~2- to 8-AU circular orbit around the main-sequence star, which is exchanged into the binary. Calculations show that there is a substantial probability (~15%) (9, 10, 13) that the planet will survive the exchange interaction and remain bound in a stable, hierarchical orbit of moderate eccentricity. The postexchange planetary orbit will expand further in response to the subsequent mass loss from the system, and we infer that the initial postexchange orbit had a semimajor axis of about 18 AU, leading to the current value of 23 AU. There will also be modest circularization of the planet's orbit during the mass-transfer phase because of the slow loss of mass from the system, so the original planet eccentricity was somewhat higher than that currently observed. Because the system is young, it has been far from the cluster core for most of the time since the exchange and for the full length of time that the planet has been in its current wide, circumbinary orbit. The stellar densities in the regions the system traversed are two to four orders of magnitude lower than in the core, and further interactions are correspondingly much less probable. Because the planetary companion is in a very wide orbit and is highly inclined to the inner binary, it is very unlikely to have formed in this orbit, for example, in mass flowing out during the neutron star spin-up phase. Hence, all the evidence supports the conclusion that the circumbinary companion of PSR B1620-26 is a few $M_{Jupiter}$ planet, originally formed around a ~0.8 to 0.9 $M_\odot$ main-sequence star in a Jupiter-like orbit.

This conclusion has important implications for understanding how the frequency of planets depends on the metal content of the parent system, with the caveat that drawing conclusions from a single observation is very uncertain. A recent transit search for planets in close orbits in the globular cluster 47 Tucanae failed to find any planets (29), leading to the conclusion that the frequency of close planets was less than observed in the solar neighborhood. Several explanations are possible; one hypothesis is simply that planets do not form in low-metallicity environments. This would be consistent with the apparent preference for metal-rich systems in the solar neighborhood (30). However, the transit search is sensitive only to those close-in planets that are believed to have migrated inwards from initially larger radii. If the unknown migration mechanism (31, 32) is dependent on metal content, then there could still be planets in AU-size orbits but not in closer orbits, because they would not migrate. Finally, it is possible that regardless of how many and where planets form, the dynamical perturbations experienced over the history of the cluster would be too disruptive to allow the survival of any planets (33–35), although if planets migrate to orbital periods as short as few days, it is very difficult to disrupt their orbits without disrupting the star also.

It appears that at least one planet did form in the globular cluster M4. M4 is a factor of 5 more metal-poor than 47 Tucanae, so the existence of a planet in M4 suggests that Jovian planets should be formed at least as efficiently in 47 Tucanae. The fact that the planet appears to have survived for most of the lifetime of the cluster in its original orbit suggests that dynamical disruption in clusters is not sufficient to completely destroy any planetary population. It appears then that the lack of close-in planets may reflect a metallicity dependence in the migration mechanism or is evidence for the crowding in young clusters to suppress migration but not formation.

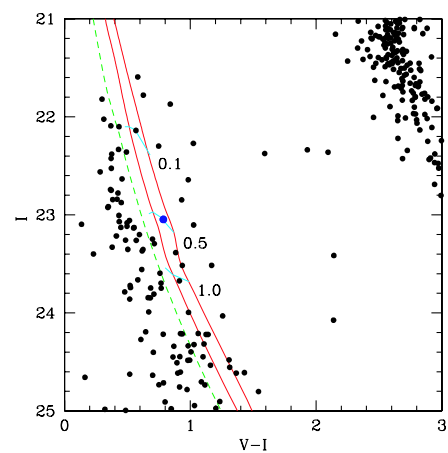

**Fig. 2.** The location of the pulsar companion in the color-magnitude diagram. The companion is plotted in blue and the rest of the cluster stars [including those from several other HST fields covering a much wider area than that discussed in (22) and that shown in Fig. 1] are in black. The green dashed line is the cooling curve for a 0.5-$M_\odot$ carbon-oxygen white dwarf (the upper envelope of the normal white-dwarf cooling sequences). The red curves are cooling curves for 0.3- and 0.4-$M_\odot$ helium-core white dwarfs with hydrogen envelopes. The three short cyan curves are isochrones (helium core only) at $0.1 \times 10^9$, $0.5 \times 10^9$, and $1.0 \times 10^9$ years, respectively. Thus, the pulsar companion is clearly undermassive and young, as expected from the evolutionary scenario.

**Table 1.** Astrometric and photometric data for M4, PSR B1620−26, and the white dwarf (WD) and red star at the edge of the error circle. The M4 coordinates are for the cluster center (38), and its proper motion is the absolute proper motion in RA and DEC, with respect to an extragalactic frame (39), where $\mu_\alpha \cos\delta$ is the projected proper motion in RA at declination $\delta$, $\mu_\delta$ is the proper motion in DEC, and both are in units of arc sec year$^{-1}$. The error in these measurements are $0.47 \times 10^{-3}$ and $0.48 \times 10^{-3}$ arc sec year$^{-1}$, respectively. The errors in the pulsar proper motion in RA and DEC are $1 \times 10^{-3}$ and $5 \times 10^{-3}$ arc sec year$^{-1}$ (5, 6). The coordinates are in the J2000 reference frame, with the pulsar position determined on JD 2,448,725.5. The stellar coordinates are also J2000 with their positions set on JD 2,449,820. The magnitudes for the two stellar objects are on the natural HST system and are not corrected for extinction. The extinctions in the F555W and F814W bandpasses are taken to be 1.31 and 0.82, respectively, and the true distance modulus to the cluster is 11.18. The stellar proper motions are also with respect to a nonmoving extragalactic reference frame. The cluster has a measured one-dimensional proper motion dispersion of about $2 \times 10^{-3}$ arc sec year$^{-1}$, which we take as our proper motion errors for the two stars discussed here. Dotted entries indicate no data.

| Object | RA | DEC | Proper motion | | F336W | F555W | F814W |
| --- | --- | --- | --- | --- | --- | --- | --- |
| | | | $\mu_\alpha \cos\delta$ | $\mu_\delta$ | | | |
| M4 | 16:23:35.5 | −26:31:31 | −12.26 | −18.95 | ... | ... | ... |
| PSR | 16:23:38.2218 | −26:31:53.769 | −13.4 | −25 | ... | ... | ... |
| WD | 16:23:38.232 | −26:31:53.364 | −10.95 | −15.28 | 23.963 | 24.093 | 23.379 |
| Red star | 16:23:38.189 | −26:31:53.040 | −10.36 | −15.54 | 24.061 | 21.437 | 19.457 |

# Atomic Memory for Correlated Photon States

C. H. van der Wal,[1] M. D. Eisaman,[1] A. André,[1] R. L. Walsworth,[2] D. F. Phillips,[2] A. S. Zibrov,[1,2,3] M. D. Lukin[1]*

We experimentally demonstrate emission of two quantum-mechanically correlated light pulses with a time delay that is coherently controlled via temporal storage of photonic states in an ensemble of rubidium atoms. The experiment is based on Raman scattering, which produces correlated pairs of spin-flipped atoms and photons, followed by coherent conversion of the atomic states into a different photon beam after a controllable delay. This resonant nonlinear optical process is a promising technique for potential applications in quantum communication.

The realization of many basic concepts in quantum information science requires the use of photons as quantum information carriers and matter (e.g., spins) as quantum memory elements (*1*). For example, intermediate memory nodes are essential for quantum communication and quantum cryptography over long photonic channels (*2*). Thus, methods to facilitate quantum state exchange between light and matter are now being actively explored (*3–5*). We report a proof-of-principle demonstration of a technique in which two correlated light pulses can be generated with a time delay that is coherently controlled via the storage of quantum photonic states in an ensemble of Rb atoms. This resonant nonlinear optical technique is an important element of a promising approach to long-distance quantum communication, proposed recently by Duan *et al.* (*6*). This proposal is based on earlier theoretical suggestions (*7, 8*) for storing photonic states in atomic ensembles [for a recent review see, e.g. (*9*)].

Our approach involves coherent control of the optical properties of an atomic ensemble and is closely related to studies involving electromagnetically induced transparency (EIT) (*10–12*) and resonantly enhanced nonlinear optical processes (*13, 14*) in a highly dispersive medium (*15, 16*). The use of such processes for nonclassical light generation has been extensively studied theoretically (*17–21*) and is probed experimentally in the present work. Storage of weak classical light pulses has been demonstrated (*22–24*); also, nonclassical spin

[1]Department of Physics, Harvard University, Cambridge, MA 02138, USA. [2]Harvard-Smithsonian Center for Astrophysics, Cambridge, MA 02138, USA. [3]P. N. Lebedev Institute of Physics, Moscow 117924, Russia.

*To whom correspondence should be addressed. E-mail: lukin@physics.harvard.edu